\documentclass[12pt]{article}
\usepackage{epsfig,graphicx,amssymb,amsmath}

\title{Neutron in Strong Magnetic Fields}
\author{M.A. Andreichikov, B.O. Kerbikov,\\
Moscow Institute of Physics and Technology, \\
Dolgoprudny, Institutskiy Pereulok 9, 141700 Moscow Region, Russia,\\
State Research Center\\Institute of Theoretical and Experimental Physics, \\
Moscow, 117218 Russia\\
\\
V.D. Orlovsky,  Yu.A. Simonov\\
State Research Center\\Institute of Theoretical and Experimental Physics, \\
Moscow, 117218 Russia}
\newcommand{\vect}[1]{\mathbf{#1}}
\newcommand{\gvect}{\boldsymbol}

\newcommand{\beq}{\begin{eqnarray}}
 \newcommand{\eeq}{\end{eqnarray}}
\newcommand{\be}{\begin{equation}}
 \newcommand{\ee}{\end{equation}}

\def\fun#1#2{\lower3.6pt\vbox{\baselineskip0pt\lineskip.9pt
\ialign{$\mathsurround=0pt#1\hfil ##\hfil$\crcr#2\crcr\sim\crcr}}}

\newcommand{{\SD}}{\rm SD}

\newcommand{\vetau}{\mbox{\boldmath${\rm \tau}$}}

\newcommand{\ver}{\mbox{\boldmath${\rm r}$}}
\newcommand{\vesig}{\mbox{\boldmath${\rm \sigma}$}}

\newcommand{\veP}{\mbox{\boldmath${\rm P}$}}

\newcommand{\veq}{\mbox{\boldmath${\rm q}$}}

\newcommand{\vez}{\mbox{\boldmath${\rm z}$}}

\newcommand{\veR}{\mbox{\boldmath${\rm R}$}}

\newcommand{\vek}{\mbox{\boldmath${\rm k}$}}
\newcommand{\ven}{\mbox{\boldmath${\rm n}$}}

\newcommand{\venu}{\mbox{\boldmath${\rm \nu}$}}
\newcommand{\vexi}{\mbox{\boldmath${\rm \xi}$}}
\newcommand{\veta}{\mbox{\boldmath${\rm \eta}$}}
\newcommand{\veB}{\mbox{\boldmath${\rm B}$}}

\newcommand{\veJ}{\mbox{\boldmath${\rm J}$}}

\newcommand{\vepi}{\mbox{\boldmath${\rm \pi}$}}

\newcommand{\lan}{\langle}
\newcommand{\ran}{\rangle}

\begin{document}
\maketitle
\begin{abstract}
Relativistic world-line Hamiltonian for strongly interacting $3q$ systems in magnetic field is derived from the path integral for the corresponding Green's function. The neutral baryon Hamiltonian in magnetic field obeys the pseudomomentum conservation and allows a  factorization of the c.m. and internal motion. The resulting expression for the baryon mass in magnetic field is written explicitly with the account of hyperfine, OPE and OGE (color Coulomb) interaction. The neutron mass is fast decreasing with magnetic field, losing $1/2$ of its value at $eB \sim 0.25$ GeV$^2$ and is nearly zero at $eB \sim 0.5$ GeV$^2$.  Possible physical consequences of the calculated mass trajectory of the neutron, $M_n(B)$, are presented and discussed.
\end{abstract}

\section{Introduction}
The properties of strongly interacting matter under extreme conditions are challenging to study both from experimental and theoretical sides. Currently a great interest attracts the response of baryon and quark matter to intense magnetic field (MF) \cite{Kharzeev:2012ph}. The outbreak of interest to this subject is caused by the fact that MF of the order of $eB\sim \Lambda^2_{QCD}\sim 10^{19}$G (GeV$^2\simeq 5.12 \cdot 10^{19}$G) became a physical reality. Such MF is created (for a short time) in peripheral heavy ion collisions at RHIC and LHC \cite{Kharzeev:2007jp}. The field about four orders of magnitude less exists on the surface of magnetars and it may be of the order of $10^{17}$G in its interior \cite{Potekhin:2011xe}. MF, as high as (100 MeV)$^2$, can change the internal structure of baryons and affect the possible neutron matter $\rightarrow$ quark matter transition, since MF can influence the phase structure of the QCD vacuum \cite{Gatto:2010pt}. Prior to analyzing the behavior of bulk neutron matter embedded in MF one should understand what happens to a neutron in MF. What are the changes that occur to its mass, shape and decay properties? Similar questions were raised before in regard to the hydrogen atom and positronium \cite{ShabadVysotsky}. In case of the hydrogen atom it was shown that in superstrong MF radiative corrections screen the Coulomb potential thus preventing the ``fall to the center'' phenomenon. As for the positronium, the collapse was predicted at super-high MF $eB\gtrsim 10^{40}$G \cite{Shabad:2007xu}.

The situation with hadron masses in presence of strong MF demands an analysis at the quark level based on the fundamental QCD principles. Quark structure comes into play when the Landau radius $r_H=(eB)^{-1/2}$ becomes equal or smaller than the size of the hadron. For example, the value of MF which corresponds to $r_H=0.6$ fm is $eB\simeq 5\cdot 10^{18}$ G. The first results obtained at the quark level have been acquired in two different approaches: on the lattice \cite{Hidaka:2012mz,Braguta:2011hq}, and analytically \cite{Andreichikov_old,Andreichikov_new,Andreichikov_asympt,Orlovsky:2013gha,Machado}. Analytical results \cite{Andreichikov_old,Andreichikov_new,Andreichikov_asympt,Orlovsky:2013gha} were obtained using the QCD path integral technique and the relativistic world-line Hamiltonian \cite{Simonov_path,path_integral_new}. Our results presented in \cite{Andreichikov_new} are in agreement with the lattice data \cite{Hidaka:2012mz,Braguta:2011hq}, in the region $eB\leq 5$ GeV$^2$, where lattice calculations in MF are reliable.

Performing the analytic calculations of meson spectra without quark loop corrections in gluon exchange in \cite{Andreichikov_old} we observed that meson mass tends to zero due to enhanced color Coulomb interaction. This phenomenon, which may be called ``The magnetic collapse in QCD'', occurs in the large $N_c$ limit, when the contribution of quark loops is negligible. Below we show that the same situation is encountered in the neutron, again in absence of quark loops. However, the inclusion of quark loop effects, done in \cite{Andreichikov_asympt}, eliminates the problem of ``magnetic collapse'' in meson, and as we show below, the same is true for baryons. Instead, one encounters in mesons the problem of the strong enhancement of the wave function at small distances, which in turn leads to the amplification of the hyperfine(hf) interaction -- the ``magnetic focusing'' effect, first found in hydrogen \cite{Andreichikov_focusing} and in any system in MF, which contain oppositely charged components \cite{Simonov_focusing}. This makes the $\pi^0$ mass at large $eB$ rather small, as it was found on the lattice \cite{Hidaka:2012mz} and in the Nambu-Goldstone type of analysis in \cite{Orlovsky:2013gha}. We show below that the neutron mass also becomes small in strong MF due to color Coulomb and hf interactions. Moreover, the first order hf contribution produce, zero neutron mass at some $B_{\rm{crit}}$, and even the smearing of the hf term, which makes meson masses  nonvanishing \cite{Andreichikov_new}, does not prevent the vanishing of the neutron mass. However, the theorem of \cite{Simonov_spin} forbids the vanishing of the mass due to MF, which implies that higher orders make this mass finite, however small.

The main result of the paper is the fast decrease of the neutron mass, what poses some questions to the dynamics of the neutron stars in strong MF and their possible transitions into quark stars.

To evaluate the baryon spectra one has to overcome several difficulties. The first problem is to develop the relativistic formalism for three particles with nonperturbative interaction. The formalism of this kind is the 3-body world-line Hamiltonian \cite{Simonov_baryon1,Simonov_baryon2}, obtained for zero MF from the Fock-Feynman-Schwinger path integral \cite{Simonov_path} and used in \cite{Simonov_spin,Simonov_IR,Simonov_Weda} for baryon spectrum. We consider this formalism in the case of three quarks in Section 2. We also show there, that in the neutral $3q$ system one can introduce pseudomomentum and exactly factorize the center of mass (c.m.) and relative motion, as it was done in the neutral 2-body system \cite{factorization}. In this way the classical factorization problem in MF, studied for decades for the neutral 2-body system, is solved here for the neutral 3-body system with arbitrary masses and charges both in nonrelativistic and relativistic context. In Section 3 we treat confinement, using for it a simplified quadratic form, which allows to find the wave function analytically with 5\% accuracy for eigenvalues, and write down the spin-flavor part of the wave function. In Section 4 we estimate the contribution of OGE (color Coulomb) interaction $\lan V_{\rm{Coul}}\ran$ for three quarks with obtained wave functions, first using gluon loop (asymptotic freedom) form, and then quark loop contribution. In Section 5 we study the spin structure of the wave function and spin splitting in MF. The situation here is similar to the spectrum of hydrogen atom or meson with hyperfine and magnetic moment interaction included. The subtle point is that the use of hf interaction proportional to the $\delta$-function at the origin in the first order perturbation theory, which results in vanishing of the neutron mass at large MF. In addition to the observed spin splitting in baryons is much stronger than in mesons, and one must introduce additional sources of the spin-spin interaction, the OPE forces, which are also subject to MF.  In Section 6 all pieces of the baryon mass are collected and results of numerical calculations for the total mass are presented as a function of MF. Section 7 is devoted to the discussion of the results and their physical significance. Concluding remarks are in Section 8 together with future prospects. Three Appendices contain the details of the calculations.

\section{Baryons in magnetic field}

Our approach to the problem of neutron properties in MF is based on recently developed theory of qurk-antiquark system in MF \cite{Andreichikov_new}. The starting point is the Feynman-Schwinger (world-line) representation of the quark Green's function. The same formalism for baryons in absence of MF was developed in \cite{Simonov_baryon1,Simonov_baryon2,Simonov_Weda} and successfully used in \cite{Simonov_Kerbikov}. Here  we accommodate the treatment of MF from \cite{Andreichikov_new} to the three-body relativistic Hamiltonian of \cite{Simonov_baryon1,Simonov_baryon2,Simonov_Weda}. Consider a neutron as a three-quark system with $d$-quarks at positions $\vez^{(1)}$ and $\vez^{(2)}$, and $u$-quark at $\vez^{(3)}$. The  relativistic free motion Hamiltonian has the form
\be \label{H0_free}
H_0 = \frac{1}{2\omega_+}\textbf{P}^2 + \frac{1}{2\omega}\vepi^2 + \frac{1}{2\omega}\textbf{q}^2 + \sum^3_{i=1} \frac{m^2_i+\omega^2_i}{2\omega_i}.
\ee
Here the momenta $\textbf{P}, \vepi$ and $\textbf{q}$ correspond to the Jacobi coordinates
\be\label{Jacobi_mom}
\textbf{P} = -i \frac{\partial}{\partial \textbf{R}}, \quad \vepi = -i \frac{\partial}{\partial \veta}, \quad \textbf{q} = -i \frac{\partial}{\partial \vexi},
\ee
where
\be \label{Jacobi_coord}
\left\{ \begin{array}{l}
\textbf{R}=\frac{1}{\omega_+} \sum \omega_i\textbf{z}^{(i)},\\
\veta = \frac{\textbf{z}^{(2)} - \textbf{z}^{(1)}}{\sqrt{2}},\\
\vexi =\sqrt{\frac{\omega_3}{2\omega_+}} (\textbf{z}^{(1)} + \textbf{z}^{(2)} -
2\textbf{z}^{(3)}).\end{array} \right.
\ee
The $i$-th quark current mass is $m_i$, the quantities $\omega_i$ play the role of constituent masses, we denote $\omega_1=\omega_2\equiv \omega, \quad \omega_u=\omega_3, \quad \omega_+ = 2\omega+\omega_3$. The momenta $\textbf{P}, \vepi$ and $\textbf{q}$ are related to the momenta of individual quarks by
\be
p_k^{(i)}= \alpha_i  P_k +\beta_i
q_k + \gamma_i \pi_k,\label{12}
\ee
\be
p_k^{(1)} = \frac{\omega}{\omega_+}P_k +\sqrt{\frac{\omega_3}{2\omega_+}} q_k-
\frac{1}{\sqrt{2}}\pi_k,\label{13}
\ee
\be
p_k^{(2)} = \frac{\omega}{\omega_+} P_k
+\sqrt{\frac{\omega_3}{2\omega_+}} q_k + \frac{1}{\sqrt{2}}\pi_k,\label{14}
\ee
\be
p_k^{(3)} = \frac{\omega_3}{\omega_+} P_k
-\sqrt{\frac{2\omega_3}{\omega_+}} q_k. \label{15}
\ee
In (\ref{H0_free}) the center-of-mass motion decouples and can be removed from the Hamiltonian.

For a neutral three-body and in general for a neutral $N$-body nonrelativistic system embedded in MF factorization of the center-of-mass motion is possible using the conserved pseudomomentum \cite{factorization, 27}. The realization of the factorization procedure depends on the relation between the masses and charges of the three particles forming the system. For the neutron $m_1=m_2=m_d$, $m_3=m_u$, $e_1=e_2=-e/2$, $e_3=e$. In strong MF we shall consider for simplicity the case of symmetrical spin configuration, when both $d$-quarks have the same spin orientation, opposite to that of $u$-quark. As will be seen, these states provide the highest and the lowest energy  eigenvalues at large $B$. For such a configuration the problem was solved in \cite{Simonov_three_body} both in the nonrelativistic and relativistic case. Below we follow the results obtained there. With MF included the Hamiltonian has the form
\be \label{2}
H_0 = \sum_{i=1}^3 \frac{(p_k^{(i)} - e_i A_k)^2 + m^2_i + \omega^2_i }{2\omega_i},
\ee
choosing the gauge $\textbf{A} = \frac{1}{2} (\textbf{B}\times \textbf{z})$ and passing to the Jacobi coordinates (\ref{Jacobi_coord}) and momenta (\ref{Jacobi_mom}) we have
\begin{multline}
 H_0= \frac{1}{2\omega} \left[ \frac{\omega}{\omega_+} \mathcal{\veP}
+\sqrt{\frac{\omega_3}{2\omega_+}}\veq -\frac{\vepi}{\sqrt{2}} +\frac{e}{4}
\left(\veB \times \left(\veR + \sqrt{\frac{\omega_3}{2\omega_+}} \vexi -
\frac{\veta}{\sqrt{2}}\right)\right)\right]^2+\\
+ \frac{1}{2\omega} \left[ \frac{\omega}{\omega_+} \mathcal{\veP}
+\sqrt{\frac{\omega_3}{2\omega_+}}\veq +\frac{\vepi}{\sqrt{2}} +\frac{e}{4}
\left(\veB \times \left(\veR + \sqrt{\frac{\omega_3}{2\omega_+}} \vexi +
\frac{\veta}{\sqrt{2}}\right)\right)\right]^2+\\
+ \frac{1}{2\omega_3} \left[ \frac{\omega_3}{\omega_+} \mathcal{\veP}
-\sqrt{\frac{2\omega_3}{\omega_+}}\veq -\frac{e}{{2}}\left(\veB \times
\left(\veR -\sqrt{\frac{2\omega^2}{ \omega_+\omega_3}} \vexi
\right)\right)\right]^2+\\
+ \sum^3_{i=1}
\frac{m^2_i+\omega^2_i}{2\omega_i} \equiv \frac{1}{2\omega}
\left((\veJ^{(1)})^2 + (\veJ^{(2)})^2\right)+\frac{1}{2\omega_3} (\veJ^{(3)})^2+
\sum^3_{i-1} \frac{m^2_i+\omega^2_i}{2\omega_i}.\label{16}
\end{multline}
The conserved pseudo-momentum for this system reads
\be
\hat{\textbf{F}} = \textbf{P} - \frac{e}{2}\sqrt{\frac{\omega_+}{2\omega_3}}(\textbf{B}\times \vexi).
\ee
The neutron wave function in MF is an eigenfunction of $\hat{\textbf{F}}$ with the eigenvalue $\textbf{F}$
\be\label{pseudo_eigen}
\hat{\textbf{F}} \Psi (\textbf{R},\vexi,\veta) = \textbf{F} \Psi (\textbf{R},\vexi,\veta).
\ee
The existence of the conserved pseudo-momentum allows to represent the wave function in the form $\Psi (\textbf{R},\vexi,\veta) = e^{i\venu \textbf{R}} \varphi(\vexi,\veta)$ and to find the phase $\venu$ from the eigenvalue equation (\ref{pseudo_eigen}). We obtain
\be\label{wave_total}
\Psi (\textbf{R},\vexi,\veta) = \exp\left\{i\left[\textbf{F} + \frac{e}{2}\sqrt{\frac{\omega_+}{2\omega_3}} (\textbf{B}\times \vexi) \right]\textbf{R} \right\} \varphi(\vexi,\veta).
\ee
Applying $J_k^{(i)} \Psi$ to the wave function (\ref{wave_total}) one gets
\be (\veJ^{(1)})^2 e^{i\venu \textbf{R}}\varphi = e^{i\venu \textbf{R}} \left[
\sqrt{\frac{\omega_3}{2\omega_+}}\left(-i\frac{\partial}{\partial\vexi}\right)
+ \textbf{C}^{(1)}\right]^2\varphi,\label{18}\ee

\be (\veJ^{(2)})^2 e^{i\venu \textbf{R}}\varphi = e^{i\venu \textbf{R}} \left[
\sqrt{\frac{\omega_3}{2\omega_+}}\left(-i\frac{\partial}{\partial\vexi}\right)
+\textbf{C}^{(2)}\right]^2\varphi,\label{19}\ee

\be (\veJ^{(3)})^2 e^{i\venu \textbf{R}}\varphi = e^{i\venu \textbf{R}} \left[
\sqrt{\frac{2\omega_3}{\omega_+}}\left(-i\frac{\partial}{\partial\vexi}\right)
-\textbf{C}^{(3)}\right]^2\varphi,\label{20}\ee
where
\be
\textbf{C}^{(1)}=\frac{\omega}{\omega_+} \textbf{F} +\frac{e}{4}
\sqrt{\frac{\omega_+}{2\omega_3} } (\veB \times \vexi) -
\frac{\vepi}{\sqrt{2}} - \frac{e}{4\sqrt{2}} (\veB\times \veta),\label{21}
\ee
\be
\textbf{C}^{(2)} = \textbf{C}^{(1)} (\vepi \to -\vepi, \veta\to -\veta),\label{22}
\ee
\be
\textbf{C}^{(3)} = \frac{\omega_3}{\omega_+} \textbf{F} + \frac{e}{4}
\sqrt{\frac{2\omega_+}{\omega_3} }(\veB\times \vexi).\label{23}
\ee
In (\ref{18})-(\ref{20}) the following combinations appear:
\be
(\veB\times \vexi)_k \frac{\partial\varphi}{i\partial \xi_k} = B_k
L_k^{(\xi)} \varphi, ~~ L_k^{(\xi)} = e_{klm} \xi_l \frac{\partial}{i\partial
\xi_m},\label{24}
\ee
\be
(\veB\times \veta)_k \frac{\partial\varphi}{i\partial \eta_k} = B_k
L_k^{(\eta)} \varphi, ~~ L_k^{(\eta)} = e_{klm} \eta_l
\frac{\partial}{i\partial \eta_m}.\label{25}
\ee
Note that the two orbital momenta $\textbf{L}^{(\xi)}$ and $\textbf{L}^{(\eta)}$ are independent and commute with each other. Finally from (\ref{16}) one obtains for $\mathbf{F} =  0$,
$$ H_0 =- \frac{1}{2\omega} (\Delta_\xi + \Delta_\eta) + \frac{1}{2\omega}
\left(\frac{eB}{4}\right)^2 \left( \frac{\omega^2_+}{\omega^2_3}
\vexi_\bot^2+ \veta_\bot^2\right)+$$ \be +\frac{e\textbf{B}}{4\omega} \left(
\frac{\omega_3 -2\omega}{\omega_3} \textbf{L}^{(\xi)} + \textbf{L}^{(\eta)}\right)+
\sum^3_{i=1} \frac{m^2_i + \omega^2_i}{2\omega_i}.\label{26}\ee

A word of caution is in order here. One can safely put $\mathbf{F} = 0$ at the ground state only when the interparticle potential is a harmonic oscillator one \cite{factorization,27,Machado}, otherwise the ground state  may require nonzero $\textbf{F}$, as it happens in the nonrelativistic treatment of heavy quarkonia \cite{Machado}.  Below we show that with the high accuracy confinement may be represented in a such form.

Next we add the interaction terms to the Hamiltonian following the approach developed in \cite{Andreichikov_new} for mesons. The complete Hamiltonian has the form
\be
H^{(B)}=H_0 +V_\sigma + V_{\rm conf} + V_{\rm Coul} + \Delta_{SE} + \Delta_{\rm string} + \Delta_{SD}.\label{1}
\ee
Here
\be
V_\sigma =- \sum^3_{i=1} \frac{e_i \vesig^{(i)} \veB}{2\omega_i}, ~~ V_{\rm conf} =\sigma \sum^3_{i=1} |\vez^{(i)}- \vez_Y|,
\label{3}
\ee
where $\vez_Y$ is the string junction position (Torricelli point),
\be
V_{\rm Coul} = - \frac{2}{3} \sum_{i>j} \frac{\alpha_s(r_{ij})}{r_{ij}}, ~~
r_{ij} \equiv |\vez^{(i)}-\vez^{(j)}|,
\label{24}
\ee
\be
\Delta_{SE} = -\frac{3\sigma}{4\pi} \sum^3_{i=1} \frac{1+\eta(\lambda(\sqrt{2eB+m_i^2}))}{\omega_i},\label{5}
\ee
where $\eta(t) = t\int_0^\infty z^2K_1(tz)e^{-z}dz$ and $\lambda\sim 1$ GeV$^{-1}$ is vacuum correlation lengths,
\be
\Delta V_{\rm string} = - \sum\frac{l^2_i\sigma \lan r_i^{-1}\ran}{2\lan
\sigma r_i\ran (\omega_i + \frac13 \lan \sigma r_i\ran)},~~
r_i=|\vez_i-\vez_Y|,\label{6}
\ee
$\Delta_{SE}$ is quark self-energy \cite{Simonov_SE}, $l_i$ is the angular momentum of the quark $i$.
The spin-dependent interaction can be splitted into four terms,
\be
\Delta_{SD} = \Delta_{ss}^{\rm pert}+ \Delta_{ss}^{\rm nonp}+
\Delta_{SO}^{\rm pert}+ \Delta_{SO}^{\rm nonp},\label{7}
\ee
and, e.g.,
\be
\Delta_{ss}^{\rm pert} = \sum_{i<j} \frac{\vesig^{(i)} \vesig^{(j)} V_4
(r_{ij}) + [3 (\vesig^{(i)} \ven) (\vesig^{(j)}\ven) -
\vesig^{(i)}\vesig^{(j)}] V_3(r_{ij})}{24 \omega_i \omega_j}\label{8}
\ee
with
\be
V_4(r) = \frac{32\pi\alpha_s}{3} \delta^{(3)} (\ver), ~~ V_3 (r) =
\frac{4\alpha_s}{r^3}.\label{9}
\ee
In what follows the tensor contribution proportional to $V_3$ in (\ref{8}) will be neglected. The reason is twofold. First, we shall be interested in lowest states with $l_i=0$. However, even in this case tensor forces may be present due to deformation of the wave function in MF, as it happens with the hydrogen atom \cite{Andreichikov_focusing}. Below it will be shown that this can occur only at $eB \gg \sigma \simeq 10^{19}$ G. Therefore the second reason to ignore $V_3$  is that this term is irrelevant at $eB\lesssim 10^{19}$ G. The term $\Delta_{ss}^{\textrm{nonp}}$ appears to be much smaller than $\Delta_{ss}^{\textrm{pert}}$ and will be neglected. For more details on spin-dependent terms in absence of MF see \cite{Simonov_baryon2}, and for the case of nonzero MF a detailed derivation is given in \cite{Simonov_spin}, where it is shown that the MF induced tensor forces are tending to zero at very large MF. There also the terms $\Delta_{ss}^{\textrm{pert}}, \Delta_{ss}^{\textrm{nonp}}$ and $\Delta_{SE}$ are derived explicitly.

\section{Simplification for lowest levels}

As in the case of mesons, we shall replace $V_{\rm conf}$ by the  quadratic
expression, which after minimization with respect to parameter $\gamma$ approaches the original form (\ref{Jacobi_coord}).
\begin{multline}
 V_{\rm{conf}} =\sigma \sum^3_{i=1} |\vez^{(i)}- \vez_Y| \to V_{\rm{conf}}^{(\gamma)} = \frac{\sigma}{2}
\left\{ \sum_{i=1}^3 \left[\frac{(\vez^{(i)}-\vez_Y)^2}{\gamma} \right] +3\gamma \right\}= \\
=3\frac{\sigma\gamma}{2} + \frac{\sigma}{2\gamma} \sum^3_{i=1} (\vez^{(i)} -
\vez_{Y})^2.\label{27}
\end{multline}
Minimization yields
\be
\min_\gamma V_{\textrm{conf}}^{(\gamma)} = \sigma\left\{ \sum^3_{i=1} (\vez^{(i)} -
\vez_{Y})^2 \right\}^{1/2} \leq \sigma \sum^3_{i=1} \left\{(\vez^{(i)}- \vez_Y)^2\right\}^{1/2}=V_{\textrm{conf}}.
\ee
We approximate the Torricelli point $\vez_Y$ by the c.m. point. This is reasonable for equal or small masses. Passing to the Jacobi coordinates we get the final expression
\be
V_{\rm conf}^{(\gamma)} = \frac{3 \sigma \gamma}{2} +
\frac{\sigma}{2\gamma} \left( \frac{\omega^2_3 + 2\omega^2}{\omega_+\omega_3}
\vexi^2 + \veta^2 \right).\label{28}
\ee
As in the case of mesons, we take the average value $\lan V_{\rm Coul}\ran$ of the OGE operator (\ref{24}) and of $\Delta^{\rm pert}_{ss}$ with the wave function $\Psi(\vexi,\veta)$, corresponding to $H_0
+ V_{\rm conf}^{(\gamma)}$. The resulting energy eigenvalue  can be considered as an upper limit for the actual energy eigenvalue. From (\ref{26}) and (\ref{28}) it is clear that this wave function factorizes, $\varphi(\vexi,\veta)=\chi(\vexi)\phi(\veta)$.

Similarly to what happens in the case of the $q\bar{q}$ system \cite{Andreichikov_new}, for $eB \gg \sigma$ our system acquires the form of an elongated ellipsoid with large axis $r_0 \approx \frac{1}{\sqrt{\sigma}}$ and
small axis $r_B = \frac{1}{\sqrt{eB}}$. This results in the increase of the Coloumb term $\lan V_{\rm Coul}\ran $ asymptotically as $\ln \left(\ln \frac{eB}{\sigma}\right)$.  As will be seen, the inclusion of quark loops in the gluon exchange stabilizes the energy of the 3-body system as in the case of mesons, discussed in \cite{Andreichikov_asympt}.

Finally $\Delta_{ss}^{\rm pert}$ is considered as a correction with the average
value $\lan \Delta_{ss}^{\rm pert}\ran $ calculated with the wave functions which
are the eigenfunctions  of the equation
\be
(H_0 + V_\sigma+V_{\rm conf}) \varphi (\veta, \vexi) =
M_0(\omega_i, \gamma) \varphi (\veta, \vexi).\label{30}
\ee
The final expression for the baryon mass is
\be
 M(B) = \bar M_0 (B) +  \Delta_{SE}
(\omega_i^{(0)}) + \lan V_{\rm Coul} (\omega_i^{(0)})\ran + \lan
\Delta_{ss}^{\rm pert}(\omega_i^{(0)})\ran,\label{31}
\ee
where $\bar M_0 (B)$ is obtained inserting into $M_0 (\omega_i, \gamma)$ the extremal values of  $\omega_i$ and $\gamma$, obtained from the
conditions
 \be
 \left. \frac{\partial  M_0 (\omega_i, \gamma)}{\partial
\omega_i}\right|_{\omega_i=\omega_i^{(0)}}=0,~~\left. \frac{\partial  M_0
(\omega_i, \gamma)}{\partial
\gamma}\right|_{\gamma=\gamma^{(0)}}=0.\label{32}
\ee
We remind, that the equation (\ref{30}) admits a separable solution $\varphi (\veta, \vexi) = \phi
(\veta) \chi (\vexi)$ with $\phi$ and $\chi$ being explicit oscillator
functions yielding the exact answer for $M_0 (\omega_i, \gamma_i)$.

The total baryon wave function can be written as
$$ \Psi_B = [\Psi^{\rm symm} (\vexi, \veta) \psi^{\rm symm} (\sigma, f) + \Psi'
(\vexi, \veta) \psi'(\sigma, f)+$$
\be
+ \Psi^{\prime\prime} (\vexi,\veta)
\psi^{\prime\prime} (\sigma, f) + \Psi^a (\vexi, \veta) \psi^a (\sigma, f) ]
\psi^a (\rm color), \label{33}
\ee
where $\psi(\sigma, f) $ is spin-flavor wave
function, while $\psi(\vexi,\veta)$ is the coordinate one; the
superscripts: symm, $a$, $\prime $, $\prime\prime$ refer to symmetric,
antisymmetric, and two-dimensional representations of 3-body permutation group;
note, that $(\vexi, \veta)$ belong to $(\prime\prime,\prime)$ representations.

We shall be interested primarily in the neutron state, and since all terms in
(\ref{33}), except for the first one,  contain nonzero  angular  momenta, hence they will be suppressed at large $B$ as
compared to the first one \cite{three_body}. Therefore we can write the combination $\psi^{\rm symm} (\sigma, f)$ for the neutron with spin down as
$$ \psi_n^{\rm symm} (\sigma, f) = \frac{\sqrt{2}}{6} \{ 2u_+ d_- d_- - d_+ u_- d_-
- u_- d_+ d_- +2d_- u_+ d_- -$$ \be -d_- d_+ u_- - d_+ d_- u_- - d_- u_- d_+ -
u_- d_- d_+ + 2 d_- d_- u_+\}.\label{34}\ee

In (\ref{34}) $u_\pm , d_\pm$ denote individual quark spin-flavor functions
with spin up or down. $\psi^{\rm symm}_n (\sigma, f)$ is normalized to unity.

The above classification is simple in absence of MF and equal
quark masses, since in this case both $H_0$ (\ref{26}) and $V_{\rm
conf}^{(\gamma)}$ (\ref{28}) are symmetric. For nonzero $\veB$ three symmetry violations occur: 1) $\veB $  violates
$O(3)(SU(2))$ symmetry and spin mixing may occur between $J=\frac12$ and
$J=\frac32$ states, 2) $\veB$ violates isospin symmetry implying mixing of
$I=\frac12$ and $\frac32$ states, 3) both  $H_0$ and $V_{\rm conf}^{(\gamma)}$
are not symmetric in quark indices for $B\neq 0$, which implies, that not all, but only some
components of Eq.(\ref{34}) are dominant ones for strong $\veB$.

Strictly speaking, when spin and isospin are not good quantum numbers, the Pauli
principle applies only to $d$-quarks in the same state. Both $H_0$ and $V_{\rm conf}^{(\gamma)}$ are symmetric with respect to
$\veta\leftrightarrow -\veta$, hence the $\phi(\eta)$ component in the wave function $\Psi (\xi,\eta) = \phi(\eta) \chi(\xi)$ has a symmetry
$\phi(\veta) = \phi(-\veta)$ and $\psi_n^{\rm symm} (\sigma, f)$ is
 symmetric in $d,d$ spin coordinates, but has no definite spin and
 isospin. The terms $-  (d_+d_- + d_- d_+) u_-$ and $d_-d_-u_+$ in (\ref{34}) meet these conditions. As will be seen, when $B$ is switched on, the neutron state gets splitted into three states (in order of growing energy): $(d_-d_-u_+)$, $(d_-d_+u_-), (d_+d_-u_-)$.

 Actually only for two combination $(d_-d_-u_+)$ and $(d_+d_+ u_-)$, our
 equations with $\omega_1 =\omega_2=\omega$ are valid, and the most general
 case with arbitrary masses and charges will be considered in the subsequent
 paper. In the present paper we consider the state $(d_-d_-u_+)$ at large MF $eB \geq \sigma$, where it is dominant for the neutron, and
in addition all other states at small MF, where pseudomomentum factorization does not hold but MF can be considered as perturbation.

\section{Mass spectrum in MF}

The solution of Eq. (\ref{30}) for the neutral 3q system in MF with
confinement, given by Eq.(\ref{28}) reduces to the solution of four independent oscillator equations. For the lowest $(d_-d_-u_+)$ state this yields
$$ \frac{M_0 (\omega_i, \gamma)}{\sqrt{\sigma}} = \Omega_{\xi\bot} +
\Omega_{\eta\bot} +\frac12 (\Omega_{\xi\parallel} + \Omega_{\eta \parallel}) +
\frac{3\sqrt{\sigma}\gamma}{2}+ $$\be+ \frac{m^2_d +\omega^2 - \frac{e}{2}
B}{\omega\sqrt{\sigma}} + \frac{ m^2_u + \omega^2_3- eB}{2 \omega_3
\sqrt{\sigma}},\label{35}
\ee
where the following notations are used,
\be
\Omega_{\xi\bot}
=\left[ \left(\frac{eB}{4\sigma}\right)^2 \frac{a^2_+}{a^2a^2_3}+ \frac{a^2_3+
2a^2}{\beta a a_+ d_3}\right]^{1/2},\label{36}
\ee
 \be \Omega_{\xi\parallel}
=\sqrt{ \frac{a^2_3+ 2a^2}{\beta a a_+ a_3}},\label{37}
\ee
\be \Omega_{\eta\bot}
=\sqrt{\left(\frac{eB}{4\sigma}\right)^2 \frac{1}{a^2}+\frac{1}{\beta a}},
\label{38}
\ee
\be \Omega_{\eta\parallel} =\frac{1}{\sqrt{\beta a}}. \label{39}
\ee
Here $\omega=a\sqrt{\sigma}, \omega_3 = a_3 \sqrt{\sigma}, \gamma=
\beta/\sqrt{\sigma},$ $ a_+ = 2a+a_3$. The resulting parameters $a, a_3, \beta$
 are to be found from the conditions (\ref{32}), which are written explicitly
in the Appendix 1. Directly from (\ref{35}) it follows that at $eB \to \infty$ $\bar M_0 = M_0
(\omega_i^{(0)}, \gamma^{(0)})$ tends to a finite limit. As for the parameters $a, a_3$ and $\beta$, they vary in the limits $1\geq a, a_3, \beta \gtrapprox 0.5$, when $eB$ grows from 0 to infinity. The mass $\bar{M}_0 (eB =0)$ for $m_q=0$ is equal to $\bar{M}_0 = 6\sqrt{\sigma}$. According to \cite{Andreichikov_new} self-energy contribution also depends on MF. For $3q$ system one has
\be
\Delta_{SE} = -2 \frac{3\sqrt{\sigma}}{4\pi a}\left[1 + \eta (\lambda \sqrt{eB + m_1^2}) \right] - \frac{3\sqrt{\sigma}}{4\pi a_3} \left[1 + \eta(\lambda \sqrt{2eB + m_3^2}) \right],
\ee
where
\be
\eta(t) = t \int_0^{\infty} z^2 K_1(tz)e^{-z} dz.
\ee
Note, that  $\Delta_{SE}$ cancels a large part of the meson mass $M_0$, which might cast a doubt on the use of $\Delta_{SE}$ as a correction. However, this approach was successfully used for the calculation of many meson and baryon masses and Regge trajectories,
for baryons see e.g \cite{Simonov_baryon1, Simonov_baryon2}, for mesons \cite{Badalian_meson}.

\begin{figure}[h]
  \centering
  \includegraphics[width=0.8\textwidth]{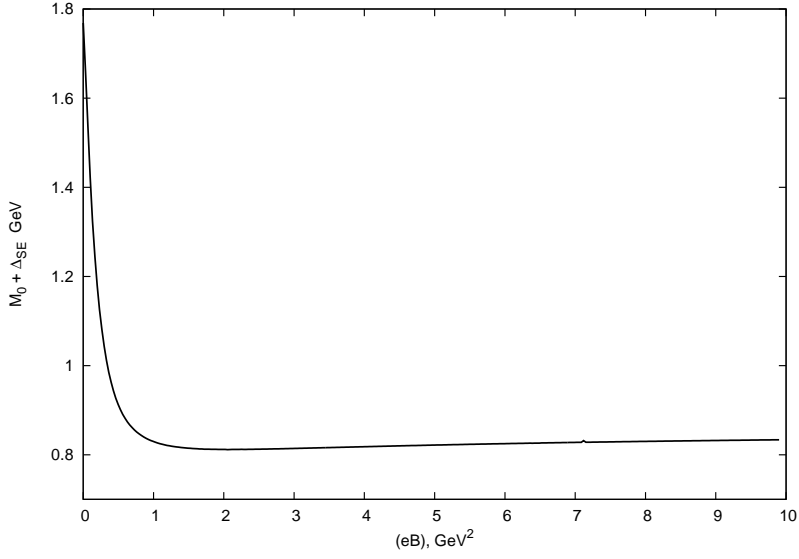}
  \caption{The dynamical baryon mass (without gluon exchange and hf interaction) in GeV as a function of $eB$.}
  \label{M0plusSE}
\end{figure}
In Fig.\ref{M0plusSE} we show the quantity $M_0 + \Delta_{SE}$ as a function of $eB$. One can see a rapid fall within the interval $0 < eB < 1$  GeV$^2$.
Consider now the color Coulomb  contribution, i.e., the term $\lan V_{\rm Coul}\ran$ with $V_{\rm Coul}$, given by (\ref{24}).

The eigenfunctions of $H_0+V_{\textrm{conf}}^{(\gamma)}$ can be written in the form
\be
\Psi(\vexi, \veta) = \psi_1 (\xi_\bot) \psi_2(\xi_\parallel) \varphi_1 (\eta_\bot)
\varphi_2(\eta_\parallel),\label{41}\ee where
$$ \psi_1 (\xi_\bot) =\frac{1}{\sqrt{\pi r^2_{\xi_\bot}}}\exp \left( -
\frac{\xi_\bot^2}{2r^2_{\xi_\bot}}\right), ~~ \psi_2 (\xi_\parallel)
=\frac{1}{(\pi r^2_{\xi_\parallel})^{1/4}}\exp \left( -
\frac{\xi_\parallel^2}{2r^2_{\xi_\parallel}}\right),$$ \be
 \varphi_1 (\eta_\bot) =\frac{1}{\sqrt{\pi r^2_{\eta_\bot}}}\exp \left( -
\frac{\eta_\bot^2}{2r^2_{\eta_\bot}}\right), ~~ \varphi_2 (\eta_\parallel)
=\frac{1}{(\pi r^2_{\eta_\parallel})^{1/4}}\exp \left( -
\frac{\eta_\parallel^2}{2r^2_{\eta_\parallel}}\right),\label{42}\ee where
$$r^{-2}_{\xi_\bot} = \omega\Omega_{\xi_\bot} \cdot \sqrt{\sigma}, ~~r^{-2}_{\xi_\parallel}=\omega
\Omega_{\xi_\parallel} \sqrt{\sigma},$$ \be r^{-2}_{\eta_\bot} =
\omega\Omega_{\eta_\bot} \cdot \sqrt{\sigma}, ~~r^{-2}_{\eta_\parallel}=\omega
\Omega_{\eta_\parallel} \sqrt{\sigma}.\label{43}\ee

Momentum space color Coulomb potential with the account of gluon and quark loop effects reads \cite{Andreichikov_asympt}
\be
V_{\rm Coul} (q) = -\frac{16 \pi \alpha_s^{(0)}}{3 \left[ q^2\left(1 + \frac{\alpha_s^{(0)}}{4\pi} \frac{11}{3} N_c \ln \left(\frac{q^2 + M_B^2}{\mu_0^2} \right)\right) + \frac{\alpha_s^{(0)}n_f |eB|}{\pi} e^{-\frac{q_{\perp}^2}{2|eB|}}T \left( \frac{q_z^2}{4\sigma}  \right)  \right]},
\label{44}
\ee
where
\be
T(z) = \frac{\ln(\sqrt{z+1} + \sqrt{z})}{\sqrt{z(z+1)}} + 1.
\ee
Inclusion of quark-antiquark loops allows to avoid an unrestricted diminishing of the mass at $eB \rightarrow \infty$. In this way the ``fall to the center'' in hydrogen atom is prevented \cite{ShabadVysotsky}. The collapse becomes a real danger only in the $N_c \rightarrow \infty$ limit.

Taking the average of the interquark OGE interaction (\ref{24}) over the wave function (\ref{41}) and keeping in mind the relation (\ref{Jacobi_coord}) between $\textbf{z}_i$ and the Jacobi coordinates, one obtains
\be
\Delta M_{\textrm{Coul}}(\rho_{\perp}(ij),\rho_z(ij)) = \int
\frac{d^2 q_{\perp} dq_z}{(2\pi)^3} V(q) e^{-\frac{q^2_\bot
\rho^2_\bot(s)}{4} -\frac{q^2_\parallel\rho^2_\parallel(s)}{4}}.\label{45}
\ee
Here
 \be \rho_\bot^2 (12) = \frac{1}{\sqrt{\left(\frac{eB}{4}\right)^2+
 \frac{a\sigma^2}{\beta}}}, ~~ \rho^2_\parallel(12) = \frac{1}{\sigma}
 \sqrt{\frac{\beta}{a}};\label{46}\ee

\be \rho_\bot^2 (13) =\rho^2_\bot (23) =
\frac{1}{\sqrt{\left(\frac{eB}{2}\right)^2+4\sigma^2
 \frac{aa_3}{\beta a^3_+}(a^2_3+2a^2)}}+\left[\left(\frac{eB}{2}\right)^2 + \frac{4\sigma^2a}{\beta }\right]^{-1/2}, \label{47}\ee
 \be \rho^2_\parallel (13) = \frac{1}{2\sigma} \left[\frac{a^3_+\beta}{a_3
 a(a_3^2+2a^2)}\right]^{1/2}+\frac{1}{2\sigma}\sqrt{\frac{\beta}{a}},\label{48}\ee
 and $\rho^2_\bot (13) =\rho^2_\bot(23),$ $\rho^2_\parallel(13)
 =\rho^2_\parallel(23)$.
 Comparing Eq.(\ref{45}) for $\lan V_{\rm Coul}\ran$ with the corresponding
 expression in case of the $(q\bar q)$ system in \cite{Andreichikov_new}, one can see the same
 structure of the integral (41) in \cite{Andreichikov_new} and our Eq. (\ref{45}), and similar values of parameters $\rho_\bot$ and
 $\rho_\parallel$, which in our case for $eB \to \infty$ behave as
 $\frac{2}{\sqrt{eB}}$ and $\frac{1}{\sqrt{\sigma}}$ respectively for $s=12$, and $\frac{2}{\sqrt{eB}}$ and $\sqrt{\frac{2}{\sigma}}$ for $s=13,23$.

This should be compared to the $(q\bar q)$ parameters $r_\bot (eB \to \infty)
=0;~ r_\parallel (eB \to \infty) = \sqrt{\frac{2}{\sigma}}$. If one represents
the color Coulomb correction for a meson as $\Delta M_{\rm Coul}^{\rm mes}
(r^2_\bot, r^2_\parallel)$, then for a baryon one can write according to
(\ref{45})
\be
 \Delta M^{\rm bar}_{\rm Coul} = \frac12 \Delta M_{\rm Coul}^{\rm
mes}(\rho^2_\bot (12), \rho^2_\parallel(12))+ \Delta M^{\rm mes}_{\rm Coul}
(\rho^2_\bot (13), \rho^2_\parallel (13)).\label{49}
\ee
Now, if one takes the standard Coulomb interaction (i.e. $V(q)$ in (48) without quark loops), we encounter the problem of boundless decrease of the neutron mass at $B\rightarrow \infty$. This phenomenon can be called the ``magnetic collapse of QCD'', which holds at least in
large $N_c$ limit when quark loop contribution becomes negligible. The situation is similar to the hydrogen atom case, where the binding energy diverges as $(-\ln^2eB)$ \cite{ShabadVysotsky}. For mesons, as it was shown in \cite{Andreichikov_old,Andreichikov_new}, $\Delta M^{\textrm{mes}}_{\textrm{Coul}}$ diverges as $-\sqrt{\sigma}\ln\ln\frac{eB}{\sigma}$ in the limit $eB \gg \sigma$. In all three cases - the hydrogen atom, mesons and baryons, the situation is cured by the screening effect produced by the loop contribution in MF. Retaining in (48) the quark loop contribution, one arrives at the nontrivial conclusion that the ground state energy is frozen and the ``fall to the center'' phenomenon is eliminated \cite{Andreichikov_asympt}. The resulting color Coulomb correction with account of screening effect from (48) is shown on Fig.2.
%We refer the reader to the Appendix ..., where it is shown that the unbounded behavior of $\Delta M^{mes}_{\textrm{Coul}}$ (and hence $\Delta %M^{bar}_{\textrm{Coul}}$) is stopped by the quark loop contribution.
\begin{figure}[h]
  \centering
  \includegraphics[width=0.8\textwidth]{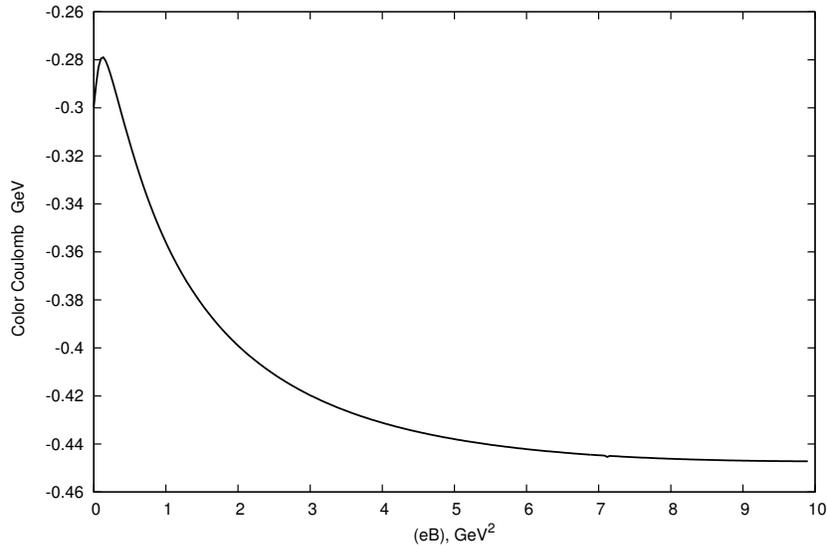}
  \caption{The color Coulomb potential contribution in GeV as a function of $eB$.One can see a saturation at $eB > 4$ GeV$^2$ due to quark loop contribution in the gluon exchange.}
\end{figure}

\section{Spin splittings in MF}

Since MF violates both spin and isospin symmetries, one must diagonalize the spin-dependent terms of the
Hamiltonian (\ref{1}) in order to find its solutions. The spin-dependent piece is
\begin{multline}
h_\sigma = \Delta^{\rm pert}_{ss} + V_\sigma= \Delta_{ss}^{\rm pert} -
\sum^3_{i=1} \frac{e_i \sigma_z^{(i)} B}{2\omega_i} \equiv \\ \equiv d\vesig_3 (\vesig_1
+\vesig_2) + b\vesig_1\vesig_2-c_3\sigma_{3z} + c(\sigma_{1z} +
\sigma_{2z}),\label{51}
\end{multline}
where
\be
 d=\frac{4\alpha_s}{9\omega\omega_3} \lan \delta
(\ver_{31})\ran, ~~ b=\frac{4\alpha_s}{9\omega^2} \lan \delta
(\ver_{12})\ran,\label{59}
\ee
\be
c=\frac{eB}{4\omega}, ~~ c_3 =\frac{eB}{2\omega_3}.\label{60}
\ee
These expressions are valid for the state $|--+\ran$. In the more general case coefficients in front of $\sigma_{1z}$ and $\sigma_{2z}$ as well as in front of $\vesig_3 \vesig_1$ and $\vesig_3\vesig_2$ should differ.

The mixing between the $S=1/2$ and $S=3/2$ states is due to the term $d\vesig_3 (\vesig_1
+\vesig_2)$. Writing the 3q spin-flavor wave function for  total
spin projection $\left( - \frac12\right)$ in a simplified form (to be
symmetrized in (123)), one has
\be
\Psi_{-\frac12} = \alpha (--+) +
\frac{\beta}{\sqrt{2}}  [ (+--)+(-+-)], ~~ \alpha^2+\beta^2=1.
\label{52}
\ee
Note, that the spin-independent part of the total Hamiltonian has a diagonal form with respect to spin variables, but diagonal elements are spin-dependent, since the quantities $\omega_i$ for the states with different spin projections are defined by a different minimization conditions. So, for the state $(--+)$ all $\omega_i$ and resulting mass $\bar{M}_0$ tend to the finite limit at large $eB$, while for the state $\frac{1}{\sqrt{2}}  [(+--)+(-+-)]$ we have one bounded and two growing $\omega_i$ at large $eB$. The resulting mass for this state grows unboundedly with increase of MF.

At zero MF  the initial values of $\alpha$ and $\beta$ are: for the neutron
$\alpha_n = \sqrt{\frac23},~~ \beta_n =-\frac{1}{\sqrt{3}}$, and for the
$\Delta$-isobar $\alpha_\Delta = \frac{1}{\sqrt{3}}, ~~ \beta_\Delta =
\sqrt{\frac23}$. Consequently one finds the ``trajectory'' of the neutron mass going down with $eB$ and that of the $\Delta$ mass going up. We shall denote these combinations $n_B$ and $\Delta_B$,
their wave functions are described by (\ref{52}) with the corresponding
$\alpha$ and $\beta$. In the limit $eB \rightarrow \infty$ we have $\alpha_n=1, \, \beta_n=0$ and $\alpha_\Delta=0, \, \beta_\Delta=1$, which corresponds to the disappearance of mixing. In the general case for a finite MF the ratio of coefficients $\beta/\alpha$ for neutron (or $\alpha/\beta$ for $\Delta$) is suppressed. Hence $|--+\rangle$ is a good approximation for lowest mass state, which gives the dominant contribution for $eB \geq \sigma$.

The trajectory $n_B$ without the hf interaction tends to a positive constant at $eB\rightarrow \infty$. The inclusion of the hf interaction at large MF can make the neutron mass negative, since $\lan V_{\rm{hf}}\ran\propto eB$. However, it was proved that MF cannot make the mass vanish due to spin-dependent forces \cite{Simonov_spin}, therefore considering the hf interaction as a perturbation, one should use the smearing factor with the smearing radius of $(0.1\div 0.2)$ fm \cite{Machado, 29,30}. As will be seen below, this procedure still does not prevent vanishing of $M_n(eB)$ at large $eB \sim 1$ GeV$^2$, which implies the importance of higher order hf interaction terms, which must ensure the positivity of $M_n(eB)$ at all values of $eB$.

The $3q$ Green's function generated by the $3q$ current $J_{\mu_1\mu_2\mu_3}$
is proportional to
$$ G\sim \lan J|n_B\ran \exp (-i M(n_B) t) \lan n_B| J\ran+$$
\be +  \lan J|\Delta_B\ran \exp (-i M(\Delta_B) t) \lan \Delta_B|
J\ran\label{55}
\ee
and therefore will display the pattern of mass oscillation
depending on MF. This is similar to the neutrino mass oscillations, but
strongly differs in scale.

\section{Baryon mass spectrum at varying MF}

In what follows we shall be interested primarily in the trajectory $n_B$ and shall use for $eB \geq \sigma$ the diagonal element of the total Hamiltonian describing the $|--+\ran$ component. The mass (energy) eigenvalue is
\be \label{En}
M_n = E+(b-2d),
\ee
where $E$ is the solution of (\ref{30}), written with account of the self-energy $\Delta_{SE}$ and the Coulomb $\lan V_{\textrm{Coul}}\ran$ corrections:
\be
E = M_{0}  + \Delta_{SE} + \lan V_{\rm Coul}\ran.\label{57}
\ee
The parameters $b$ and $d$ are defined in (\ref{59}), (\ref{60}), the explicit expressions for $\lan \delta(\ver_{ij})\ran$ are given in the Appendix 2.

\begin{figure}[h]
  \centering
  \includegraphics[width=0.8\textwidth]{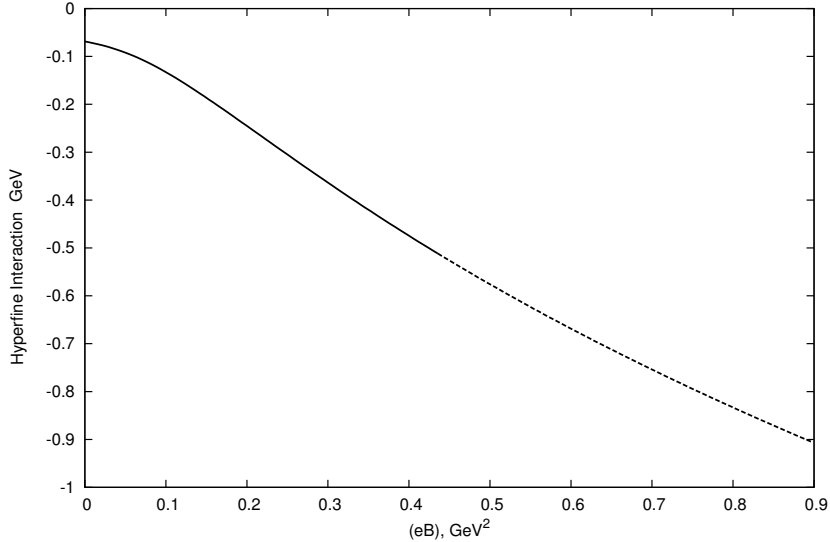}
  \caption{The hyperfine diagonal contribution $\langle V_{hf} \rangle = \tilde{b} - 2\tilde{d}$ from Eq.(\ref{vhf}) to the neutron mass in GeV as a function of $eB$.}
\end{figure}

The quantities $V_{\rm Coul} $ and $\Delta_{SE}$ are evaluated making use of the
variational averaging  procedure, hence one should find the stationary value of $M_0$ from the
conditions (\ref{32}), where $M_0$ is given in (\ref{35}).

As a result one obtains $\bar M_0 = M_0 (\omega^{(0)}, \omega_3^{(0)},
\gamma^{(0)})$, with parameters taken at stationary points. In this way $\bar
M_0(B)$ is obtained. The starting point is $eB =0$, where one has from Appendix 3 (expression (\ref{En}) is not a good approximation for zero MF)
\be
M_\pm = E + b-d \pm 3d,\label{61}
\ee
so the $n-\Delta$ mass difference is $6d \cong 0.15 \alpha_s \sqrt{\sigma} \approx 20$ MeV for $\alpha_s=0.35$ and $6d \approx 100$ MeV for $\alpha_s=1.72$.

Thus we see, that $\Delta_{ss}^{\textrm{pert}}$ by itself does not ensure the experimental splitting
between $n$ and $\Delta$ close to 300 MeV. As it is well known \cite{Simonov_Weda}, this
difference can be explained adding the OPE  interaction, having the same $\vesig_i
\vesig_j$ structure. Therefore one has to include the OPE quark-antiquark interaction
\be V_{\rm ope}^{(ij)} (\vek) = 4 \pi
g^2_{qq\pi} \vetau (i) \vetau(j) \frac{\Gamma_i\Gamma_j}{\vek^2 + m^2_\pi }
\left(\frac{\Lambda^2}{\Lambda^2+\vek^2}\right)^2,\label{62}
\ee
where $\Gamma_i =\frac{\vesig(i) \vek}{\omega_i + m_i}, ~~ \omega=\sqrt{\vek^2+m^2_i}$.
Comparing $V_{\textrm{ope}}$ (\ref{62}) with $\Delta_{ss}$ (\ref{59}), one can see that
both have the similar structure in the $p$-space, since for vanishing masses
$m_u =m_d=m_\pi=0$ one has in (\ref{62}) the structure $\frac{(\vesig(i)
\vek)(\vesig(j)\vek)}{\omega_i\omega_jk^2} \to \frac{\vesig(i)
\vesig(j)}{\omega_i \omega_j}.$ Numerically, as shown in \cite{Simonov_Weda} for $\sigma =0.12$
GeV$^2$ the contribution of $\bar V_{ss} = \sum_{i>j} (V_{hf}^{(ij)}+
V_{\textrm{ope}}^{(ij)})$ to $n$ and $\Delta$ masses are (-471 MeV) and (-79 MeV)
respectively. Therefore after summing $\Delta_{ss}^{\textrm{pert}}$, Eq.(55) and $V_{\textrm{ope}}$, Eq.(63), we introduce the new hf interaction
\be
V_{hf} = \Delta_{ss}^{\rm pert} + V_{\rm ope} \simeq \tilde{d}\sigma_3(\sigma_1 + \sigma_2) + \tilde{b}\sigma_1\sigma_2,
\label{vhf}
\ee
where the form (56) with $\alpha_s$ replaced by $\alpha_{hf} = \alpha_s + \alpha_{\textrm{ope}}$, and $\alpha_{\textrm{ope}}$ takes into account the pion charge structure of Eq.(67), see Appendix 3 for details.

\begin{figure}[h]
  \centering
  \includegraphics[width=0.8\textwidth]{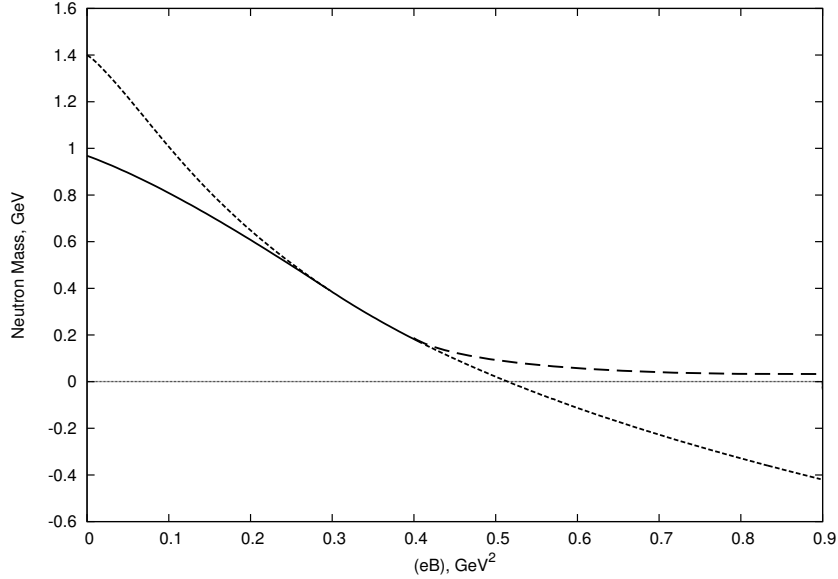}
    \caption{The neutron mass with hf correction included vs $eB$, the solid line: refers to the region where approximation made are reliable. The dotted line refers to the state $|--+\rangle$ with hf as a perturbation. Dashed line shows a possible form of the behaviour, satisfying the stabilization theorem}
\end{figure}

The difficulty we encounter here is that in order to get a correct answer it is necessary to take into account the mixing of different spin states at $eB \ll \sigma$ (see (\ref{52})). While keeping only the state $|--+\rangle$ the neutron mass at $eB \ll \sigma$ exceeds the experimental value.

Consider now the OPE interaction at growing $eB$. We have to split the OPE interaction into the contributions from $\pi^+, \pi^-$ and $\pi^0$ mesons.
\begin{multline}
V_{\textrm{ope}}^{ij} = \frac{4\pi g^2}{\omega_i \omega_j} \left[ \frac{(\gvect{\sigma}_i \cdot \vect{k})(\gvect{\sigma}_j \cdot \vect{k})}{k^2 + m_{\pi^+}^2}2 \tau_+^i \tau_-^j + \frac{(\gvect{\sigma}_i \cdot \vect{k})(\gvect{\sigma}_j \cdot \vect{k})}{k^2 + m_{\pi^-}^2}2 \tau_-^i \tau_+^j + \right.\\
\left. \frac{(\gvect{\sigma}_i \cdot \vect{k})(\gvect{\sigma}_j \cdot \vect{k})}{k^2 + m_{\pi^0}^2}\tau_3^i \tau_3^j \right]\left( \frac{\Lambda^2}{k^2 + \Lambda^2}  \right)^2.
\label{ope}
\end{multline}
As it was shown analytically in \cite{Orlovsky:2013gha} and on the lattice \cite{Bali_Foder}, the $\pi^{\pm}$ masses grow with MF as $\sim \sqrt{eB}$. Therefore the first two terms in (\ref{ope}) are suppressed at large $eB$. On the other hand, the mass of $\pi_0$ becomes somewhat smaller \cite{Orlovsky:2013gha} and its contribution into $V_{\textrm{ope}}$ important in the whole interval of MF. Hence only the last term in (67) survives in the large $eB$ limit. That's why $\alpha_{\textrm{ope}}(eB \gg \sigma) \simeq \frac{1}{3}\alpha_{\textrm{ope}}(B=0)$. Being averaged over the $|ddu \rangle$ isospin state and over the wave function (\ref{41}), the OPE and spin-spin interaction operators have the same structure with the only qualitative difference concerning the smearing procedure of the $\delta$-function - the gaussian one for the spin-spin \cite{Andreichikov_new} and the Yukawa form-factor for OPE. This difference is of a minor importance for the dependence of the interaction on MF. Therefore both corrections can be treated in a uniform way by the introduction of the effective hyperfine interaction constant $\alpha_{hf}$. Here one must distinguish two regions 1) $eB \leq \sigma$, 2) $eB \geq \sigma$. In the first one must keep all terms of the wave function as in (58), and calculate the ground state of $h_{\sigma}$ (55), as shown in Appendix 3. Here $\alpha_{hf}$ is chosen to reproduce the $\Delta-n$ splitting of $300$ MeV. In the second region one keeps only the dominant $|--+\rangle$ state, and uses Eq.(60) to calculate $M_n(B)$, the exact procedure and numerical values are given in Appendix 3.  The result is shown in Fig.3. The main general conlcusion is that the spin-spin interaction is extremely sensitive to MF. Due to the Fermi-Breit $\delta$-type interaction the mass tends to cross the $M=0$ value at $eB \sim 2\sigma$, while the general statement (see \cite{Simonov_spin}) forbids this happen.
This means that for $\delta$-type interactions the perturbation theory fails to lead to physically correct results in the limit of strong MF. One has to develop an alternative approach to treat hyperfine interaction in MF.

\section{Discussion of the results}

At this point one must look more closely at the problem of the hyperfine interaction in baryons. It was understood rather early (see e.g. the discussion in \cite{Isgur,Narodetskii}), that the standard hf interaction is too weak for a reasonable $\alpha_s$ to explain the $300$ MeV splitting between the masses of $\Delta$ and nucleon. This is contrast to the meson case, when the $q\bar{q}$ hf interaction yields $\simeq 200$ MeV splitting of $\rho$ and $\pi$ masses, and in addition the Nambu-Goldstone mechanism shifts the pion mass to its proper place. In baryon with $\alpha_S(m_N) \sim 0.5$ one obtains the splitting around $35$ MeV instead of $300$ MeV. To save the situation in \cite{Isgur} the authors have used the smeared form of the hf potential to all orders, which strongly enhanced the hf contribution: as shown in \cite{Narodetskii} for the smearing parameter $\lambda = 1.5$ GeV the hf splitting grows approximately 10 times, when taken to all orders of the $V_{hf}$, derived in the first order of $\alpha_s$.

Another approach was used in \cite{Loering1}, where the instanton interaction was parametrised to increase the hf  contribution.

Instead we have used a more physical mechanism, which should anyhow be present in the $3q$ system: the pion exchange. We have shown based on earlier papers \cite{Simonov_Weda}, that the pion exchange strongly increases the resulting gap in masses and can ensure the physical splitting (in absence of MF) for reasonable values of pion coupling with quarks.

However for growing MF one encounters several difficulties. First of all, the pseudomomentum factorization (\ref{12}), which is the basis of our present approach, requires equality of masses and energies $m_1 = m_2,\ \omega_1 = \omega_2$, which is true only for the state $|ddu\rangle \otimes |--+\rangle$. Now this state is the main component of the ground state for $eB \gg \sigma$, and therefore one obtains a reliable result for the neutron mass in this region before the inclusion of the hf interaction. However, including the hf interaction with the pion exchange at strong MF one immediately obtains a huge shift down of the neutron mass, making it negative around $eB \sim 0.5$ GeV$^2$.

This happens both with or without the pion exchange term, provided the starting $\Delta-n$ splitting is around $300$ MeV, and the problem is that the resulting hf shift at the perturbative level is huge, and violates the theorem of \cite{Simonov_spin}, stating, that MF cannot make the hadron mass to become negative. As shown in \cite{Simonov_spin}, when mass tends to zero in MF, the Dirac eigenvalues of all quarks can condense near the zero point, similarly to the case of the chiral symmetry breaking phenomenon, and may ensure the mass to be nonzero, however small. We illustrate this behavior in Fig.4 by a dashed line, which gives the idea of true trajectory, satisfying the stabilization theorem of \cite{Simonov_spin}.

At small MF we have another difficulty - inapplicability of the pseudomomentum factorization (\ref{12}), when all components of the wave function are taken into account, and to proceed, we have used the limit of small MF and calculated the neutron mass up to the order $\left( \frac{eB}{\sigma} \right)^2$, (polarizability region), using all components of the wave function. This result, valid for $eB < 0.15$ GeV$^2$, is shown in Fig.4 by a piece of a solid line below the dotted line, the latter depicts the mass of the state $|--+\rangle$, continued to the region of small $eB$, where it is not reliable. The regime of the strong MF ($eB > \sigma$), is depicted by a dotted line in Fig.4. Thus the pseudomomentum factorization method with the $|--+\rangle$ component provides the results, shown in Fig.4 by a dotted line. At larger $eB$ one assumes the saturating behavior, shown by a dashed line, while the dotted line describes the behavior predicted by the first order perturbation theory. Thus the solid line in Fig.4 shows the results obtained within the reliable approximations.

\section{Conclusions}

In our treatment of the relativistic $3q$ system embedded in MF we relied on pseudomomentum factorization of the wave function and the relativistic Hamiltonian technique. To our knowledge this is the first investigation of the three-body system with relativistic interaction in the external MF. The focus was on the dependence of the neutron mass on MF. This problem was solved analytically with confinement, color Coulomb and spin-spin interactions taken into account. From the physical arguments it is clear that MF starts to produce drastic variation of the neutron mass as soon as its strength approaches the string tension, $eB \sim \sigma \sim 10^{19} \ G \sim 0.2$ GeV$^2$. Our calculations confirm this conclusion. In strong MF the ground state of $ddu$ system has the spin structure $|--+\rangle$. An intriguing question is whether the mass of this state goes to zero in the limit $eB \rightarrow \infty$. This ''fall to the center'' phenomenon might happen for two reasons. The first one is the color Coulomb interaction. This kind of collapse is avoided due to quark-antiquark loops in the same way, as it happens in quark-antiquark system, or in the hydrogen atom due to $e^+ e^-$ loops. The second potential source of collapse is the spin-spin interaction which is proportional to the delta-function and gives a contribution  growing linearly with $eB$. How to treat this interaction beyond the perturbation theory is an old and still unresolved problem. The standard way to overcome this difficulty is to smear a delta-function around the origin with some characteristic range. For the quark system this range is given by the correlation length of the gluon field equal to $0.1-0.2$ fm. However, even with smearing the neutron mass can become zero at a finite value of $eB$ and, as it shown in \cite{Simonov_spin} this cannot happen for any value of $eB$ in the exact treatment, and the mass vanishing is the result of unlawful use of perturbation theory. Instead, the condensation of the quasi-zero Dirac eigenmodes may prevent this type of collapse.

In future study this line of research can be continued in several directions.
% First, to calculate the neutron polarizablity in arbitrary MF. At present the value of the neutron polarizability is rather uncertain both from %experimental and theoretical sides\cite{alpha, beta, gamma}.
Our method allow to consider the phase transition between neutron and quark matter in MF. This problem is of outmost importance for the neutron stars physics.

\section{Acknowledgements}

The authors are thankful to M.I. Vysotsky, S.I. Godunov and A.E. Shabad for remarks and discussions.

\newpage

\appendix
\renewcommand{\thesection}{Appendix~\arabic{section}}
\section{Solution of the system of equations (\ref{32})}

%\vspace{2cm}
 \setcounter{equation}{0}
\renewcommand{\theequation}{1.\arabic{equation}}

In terms of $a, a_3, \beta$ the Eqs. (\ref{32}) can be written as
\be\label{mimimum1}
\frac{\partial}{\partial a}\left(\Omega_{\xi\bot} +
\Omega_{\eta\bot} +\frac12 \Omega_{\xi\parallel} + \frac12 \Omega_{\eta \parallel} \right) - \frac{m_d^2-\frac{e}{2}B}{a^2\sigma} +1 =0,
\ee
\be\label{minimum2}
\frac{\partial}{\partial a_3}\left(\Omega_{\xi\bot} +
\Omega_{\eta\bot} +\frac12 \Omega_{\xi\parallel} + \frac12 \Omega_{\eta \parallel} \right) - \frac{m_u^2-eB}{2a_3^2\sigma} +\frac12 =0,
\ee
\be\label{minimum3}
\frac{\partial}{\partial \beta}\left(\Omega_{\xi\bot} +
\Omega_{\eta\bot} +\frac12 \Omega_{\xi\parallel} + \frac12 \Omega_{\eta \parallel} \right) +\frac32 =0.
\ee
Using (\ref{36})-(\ref{39}), one can calculate all terms in (\ref{mimimum1})-(\ref{minimum3}). We shall explicitly write down the results in two opposite limits: $eB=0$ and $eB\rightarrow \infty$.
\begin{description}
  \item[a)]  $eB=0$. In this case $a_3=a$ and, neglecting quark masses $m_u, m_d$, one has
  \be
  \frac{M_0}{\sqrt{\sigma}} = \frac{3}{\sqrt{\beta a}} + \frac32 (a+\beta).
  \ee
  Eq. (\ref{mimimum1}) yields $a=\beta^{-1/3}$. From (\ref{minimum3}) one has $\beta = a^{-1/3}$, which results in $a(eB=0) = \beta(eB=0)=1$.
  \item[b)] $eB \rightarrow \infty$. In this case (\ref{mimimum1})-(\ref{minimum3}) yield correspondingly
    \be\label{min1}
    4a^{3/2}\beta^{1/2} = 1+ \frac{\sqrt x(x^2+4x-2)}{(2+x)^{3/2}(2+x^2)^{1/2}},
    \ee
    \be\label{min2}
    a^{3/2}\beta^{1/2} = \frac{2+2x-x^2}{(2+x)^{3/2}x^{3/2}(2+x^2)^{1/2}},\\
    \ee
    \be\label{min3}
    6a^{1/2}\beta^{3/2} = 1 + \sqrt{\frac{x^2+2}{x(x+2)}},
    \ee
 where $x\equiv \frac{a_3}{a}$. Numerical solution of (\ref{min1})-(\ref{min3}) yields $x=1, \, \beta/a = 1$, and finally one obtains
 \be
 a(eB \rightarrow \infty) = a_3(eB \rightarrow \infty) = \beta(eB \rightarrow \infty) = \frac{1}{\sqrt{3}}.
 \ee
\end{description}

\section{Hyperfine matrix elements}

\setcounter{equation}{0}
\renewcommand{\theequation}{2.\arabic{equation}}

To calculate $\lan\delta(\textbf{r}_{13})\ran$ one can use wave functions (\ref{41})-(\ref{43}) and the relations
\be\label{a3.1}
\ver_{13} \equiv \vez_1-\vez_3 = \frac{1}{2}\left(\sqrt{\frac{2\omega_+}{\omega_3}}\vexi-\sqrt{2}\veta\right), \, \ver_{12} = \sqrt{2}\veta,
\ee
which yields
\be\label{a3.2}
\lan\delta(\textbf{r}_{13})\ran = 2^{3/2}\int\psi_1^2(\vexi_\perp)\psi_2^2(\xi_\parallel) \varphi_1^2\left(\sqrt{\frac{\omega_+}{\omega_3}}\vexi_\perp\right) \varphi_2^2\left(\sqrt{\frac{\omega_+}{\omega_3}}\xi_\parallel\right) d^3\xi.
\ee
Inserting in (\ref{a3.2}) the explicit expressions (\ref{42}), (\ref{43}), one has
\be\label{a3.3}
\lan\delta(\textbf{r}_{13})\ran = \left(\frac{2a\sigma}{\pi} \right)^{3/2} \frac{\Omega_{\xi_\perp} \Omega_{\eta_\perp}}{\Omega_{\xi_\perp}+\frac{\omega_+}{\omega_3}\Omega_{\eta_\perp}} \left[ \frac{\Omega_{\xi_\parallel} \Omega_{\eta_\parallel}}{\Omega_{\xi_\parallel}+\frac{\omega_+}{\omega_3}\Omega_{\eta_\parallel}}\right]^{1/2},
\ee
\be\label{a3.4}
\lan\delta(\textbf{r}_{12})\ran = \left(\frac{a\sigma}{2\pi} \right)^{3/2} \Omega_{\eta_\perp} \Omega_{\eta_\parallel}^{1/2}.
\ee

Now we replace $\delta(\textbf{r})$, for which the perturbation theory is unlawfull, by a smeared out version
\be
\delta^{(3)}(\textbf{r}) \rightarrow \tilde{\delta}^{(3)}(\textbf{r}) = \left( \frac{1}{\lambda\sqrt{\pi}}\right)^3 e^{-\textbf{r}^2/\lambda^2}, \quad \lambda\sim 1~\textrm{GeV}^{-1}.
\ee
With this function we obtain
\begin{multline}\label{delta13}
\lan\tilde{\delta}^{(3)}(\textbf{r}_{13})\ran = \left( \frac{2a\sigma}{\pi}\right)^{3/2}  \left[1 + \frac{2\lambda^2 a_3}{a_+}a\sigma\Omega_{\xi_\perp} \right]^{-1}  \left[1 + \frac{2\lambda^2 a_3}{a_+}a\sigma\Omega_{\xi_\parallel} \right]^{-1/2} \times \\
\Omega_{\xi_\perp} \Omega_{\eta_\perp} \Omega_{\xi_\parallel}^{1/2} \Omega_{\eta_\parallel}^{1/2} \left[ \frac{a_+}{a_3}\Omega_{\eta_\perp} + \frac{\Omega_{\xi_\perp}}{1+\frac{2\lambda^2 a_3}{a_+}a\sigma \Omega_{\xi_\perp}}\right]^{-1} \left[ \frac{a_+}{a_3}\Omega_{\eta_\parallel} + \frac{\Omega_{\xi_\parallel}}{1+\frac{2\lambda^2 a_3}{a_+}a\sigma \Omega_{\xi_\parallel}}\right]^{-1/2},
\end{multline}
\be\label{delta12}
\lan\tilde{\delta}^{(3)}(\textbf{r}_{12})\ran = \left( \frac{a\sigma}{\pi}\right)^{3/2} \Omega_{\eta_\perp} \Omega_{\eta_\parallel}^{1/2} \frac{1}{2+\lambda^2 a \sigma \Omega_{\eta_\perp}} \frac{1}{\sqrt{2+\lambda^2a \sigma \Omega_{\eta_\parallel}}}.
\ee
Eqs. (\ref{36})-(\ref{39}) help to express the r.h.s. of (\ref{delta13}), (\ref{delta12}) in terms of $a_3, a, \beta$.

\section{Baryon mass in weak MF}

\setcounter{equation}{0}
\renewcommand{\theequation}{3.\arabic{equation}}

Calculation of the mass spectrum of the 3q system in weak MF in our formalism is similar to the calculation of the Zeeman splitting in ordinary quantum mechanics. First of all, one should fix values of $a_0 = a(B=0)$, $a_{30}=a_3(B=0)$ and $\gamma_0=\gamma(B=0)$ for the zero MF, i.e. we exclude any influence of the MF over the dynamics and spatial wave function. The next step is to treat magnetic moments and hyperfine terms as a perturbation around the $E_0 = E(B=0)$ from Eq.(61). The third step is to diagonalize the
spin-dependent Hamiltonian (55) (with $\tilde{d}$ and $\tilde{b}$ from (64)) with respect to the 3q spin-flavor wave function with total spin projection $\left(-\frac{1}{2}\right)$
\be
h_\sigma = \tilde{d}\vesig_3 (\vesig_1
+\vesig_2) + \tilde{b}\vesig_1\vesig_2-c_3\sigma_{3z} + c(\sigma_{1z} +
\sigma_{2z}),
\ee
\be
\Psi_{-\frac12} = \alpha (--+) +
\frac{\beta}{\sqrt{2}}  [ (+--)+(-+-)], ~~ \alpha^2+\beta^2=1.
\ee
After straightworward manipulations one has for $n$ and $\Delta^0$
\be
M_{\pm} = E_0 + \tilde{b} - \tilde{d} - c \pm \sqrt{8\tilde{d}^2 + (c + c_3 + \tilde{d})^2},
\ee

The final step is to choose an appropriate $\alpha_{hf} = \alpha_{s} + \alpha_{\textrm{ope}}$ constant. There are three key points the choice is based on: first of all the hf interaction should provide the proper value of the splitting between the $n$ and $\Delta^0$ at zero MF, this requirement gives us $\alpha_{hf}(B=0)=17$. The second point is that in high MF limit $\alpha_{\textrm{ope}}(eB \gg \sigma) \simeq \frac{1}{3}\alpha_{\textrm{ope}}(B=0)$ since only $\pi^0$ contribution survives at high $eB$. The third point is that in the intermediate region near the $eB \sim \sigma$ these two trajectories should have a smooth connection, which provides $\alpha_{hf}(eB \gg \sigma) = 7$. This situation takes place only if $\alpha_{s}=2$ and $\alpha_{\textrm{ope}}(B=0) = 3\alpha_{\textrm{ope}}(eB \gg \sigma) = 15$.

\end{document}